\documentclass[aps,prd,twocolumn,superscriptaddress]{revtex4-1}
\usepackage{graphicx}
\usepackage{color}
\usepackage{amsfonts}
\usepackage{amsmath}

\newif\ifoneauthor
\oneauthortrue

\DeclareMathAlphabet{\mathpzc}{OT1}{pzc}{m}{it}
\definecolor{gray}{gray}{0.4}

\begin{document}

\title{A Direct Approach for the Fluctuation-Dissipation Theorem\\ under Non-Equilibrium Steady-State Conditions}

\author{Kentaro Komori}
\email[]{komori@granite.phys.s.u-tokyo.ac.jp}
\affiliation{Department of Physics, University of Tokyo, Bunkyo, Tokyo 113-0033, Japan}
\author{Yutaro Enomoto}
\affiliation{Department of Physics, University of Tokyo, Bunkyo, Tokyo 113-0033, Japan}
\author{Hiroki Takeda}
\affiliation{Department of Physics, University of Tokyo, Bunkyo, Tokyo 113-0033, Japan}
\author{Yuta Michimura}
\email[]{michimura@granite.phys.s.u-tokyo.ac.jp}
\affiliation{Department of Physics, University of Tokyo, Bunkyo, Tokyo 113-0033, Japan}
\author{Kentaro Somiya}
\affiliation{Department of Physics, Tokyo Institute of Technology, Meguro, Tokyo 152-8550, Japan}
\author{Masaki Ando}
\affiliation{Department of Physics, University of Tokyo, Bunkyo, Tokyo 113-0033, Japan}
\author{Stefan W. Ballmer}
\email[]{sballmer@syr.edu}
\affiliation{Department of Physics, Syracuse University, NY 13244, USA}
\affiliation{Department of Physics, University of Tokyo, Bunkyo, Tokyo 113-0033, Japan}

\date{\today}

\begin{abstract}
The test mass suspensions of cryogenic gravitational-wave detectors such as the KAGRA project are tasked with extracting the heat deposited on the optics. Thus these suspensions have a non-uniform temperature, requiring the calculation of thermal noise in non-equilibrium conditions. While it is not possible to describe the whole suspension system with one temperature, the local temperature anywhere in the system is still well defined. We therefore generalize the application of the fluctuation-dissipation theorem to mechanical systems, pioneered by Saulson and Levin, to non-equilibrium conditions in which a temperature can only be defined locally. The result is intuitive in the sense that the temperature-averaging relevant for the thermal noise in the observed degree of freedom is given by averaging the temperature field, weighted by the dissipation density associated with that particular degree of freedom. After proving this theorem we apply the result to examples of increasing complexity: a simple spring, the bending of a pendulum suspension fiber, as well as a model of the KAGRA cryogenic suspension. We conclude by outlining the application to non-equilibrium thermo-elastic noise.
\end{abstract}

\pacs{42.79.Bh, 95.55.Ym, 04.80.Nn, 05.40.Ca}

\maketitle

\section{Introduction}
State of the art gravitational wave detectors like KAGRA \cite{PhysRevD.88.043007}, Virgo \cite{0264-9381-32-2-024001} and Advanced LIGO \cite{PhysRevLett.116.131103} are limited by various types of thermal noise across a large fraction of their observation band. Of particular interest in this context are coating thermal noise of the test masses and thermal noise of the suspension system \cite{Harry:06,MATICHARD2015273,MATICHARD2015287,BRACCINI2005557,ACERNESE2010182}. According to the fluctuation-dissipation theorem \cite{PhysRev.83.34,0034-4885-29-1-306}, any form of energy dissipation will lead to an associated noise source: Brownian thermal noise for mechanical dissipation, and thermo-elastic noise for diffusion losses. Saulson \cite{PhysRevD.42.2437}, Levin \cite{PhysRevD.57.659,LEVIN20081941} and others \cite{PhysRevD.78.102003,PhysRevD.91.023010} applied the fluctuation-dissipation theorem to mechanical systems. They however assumed the whole system is in thermal equilibrium.

KAGRA, as well as some concepts of future gravitational wave observatories \cite{ETDesign,0264-9381-27-19-194002,PhysRevD.91.082001,0264-9381-34-4-044001}, plans to use cryogenic cooling of the test masses to further reduce the thermal noise. This however means that the suspension system - additionally tasked with heat extraction - no longer can be described by a single temperature. As a result, some confusion has arisen in the community on how exactly to describe thermal noise when temperature is only locally defined. In this paper we attempt to clear up this confusion, and give an explicit form of the fluctuation-dissipation theorem valid for non-equilibrium, but stationary conditions.

We start with the fluctuation-dissipation theorem in its various forms in section \ref{thm}. We then continue with discussing its limitations (section \ref{LIM}), and deriving the theorem (section \ref{sec:derivation}).
Next we then look at various systems of increased complexity: a simple suspension fiber, and the KAGRA suspension model where Brownian thermal noise is dominant. Finally, we outline the application to thermo-elastic noise (section \ref{sec:THELN}) and conclude.

\section{The Theorem}
\label{thm}
The most practical form of the fluctuation theorem goes back to Levin's papers \cite{PhysRevD.57.659, LEVIN20081941}. They explicitly relate the thermal noise at a given frequency seen by a readout mechanism such as an interferometer to the loss seen in the associated degree of freedom when driven at the given frequency. Both papers however assume one equilibrium temperature for the whole system. We show in this paper that for a system that is only in local thermal equilibrium this theorem can be generalized as described in following paragraphs.

If we are interested in the thermal noise associated with the degree of freedom $x=\int {q}(\vec{r}){\bf y}(\vec{r}) d^3r $, where ${\bf y}(\vec{r})$ describes the microscopic displacement of the system, and ${q}(\vec{r})$ are readout weights, then
the single-sided (i.e. only positive frequencies) displacement power spectral density $S^1_{x x}$ of this degree of freedom $x$ is given by
\begin{equation}
\label{eq:MasterNoise}
S^1_{x x}(f) = \frac{8 k_B}{\omega^2} \int d^3r \frac{w_{\rm diss}(\vec{r},f)}{F_0^2} T(\vec{r}) \, ,
\end{equation}
where $T(\vec{r})$ is the stationary temperature profile, $\omega=2 \pi f$ is the angular frequency, and $w_{\rm diss}(\vec{r},f)$ is the power dissipation density associated with driving the system with the external (generalized) force profile 
\begin{equation}
\label{eq:MasterNoiseDrive}
F(\vec{r},t) = F_0 {q}(\vec{r}) \cos(2 \pi f t)).
\end{equation}
Here $F_0$ is an arbitrary normalization of the drive amplitude that drops out in Eq.\ (\ref{eq:MasterNoise}).
This version of the Fluctuation-Dissipation theorem is identical to the one in \cite{PhysRevD.57.659}, except that the temperature profile $T(\vec{r})$  is inside the dissipation integral. In this form the theorem is applicable to any form of thermal noise, that is in particular Brownian thermal noise due to mechanical loss, as well as thermo-elastic noise due to heat diffusion. In the latter case the generalized driving force is a driving entropy, or heat load.

For a mechanical loss the dissipation density $w_{\rm diss}(\vec{r},f)$ can be rewritten in terms of loss angle $\phi(\vec{r},f)$ and maximal elastic energy density $u_{\rm max}(\vec{r},f)$ via $w_{\rm diss}(\vec{r},f)= \omega u_{\rm max}(\vec{r},f) \phi(\vec{r},f)$, resulting in
\begin{equation}
\label{eq:MasterNoiseMech}
S^1_{x x}(f) = \frac{8 k_B}{\omega} \int d^3r \frac{u_{\rm max}(\vec{r},f)}{F_0^2} \phi(\vec{r},f) T(\vec{r}).
\end{equation}

While this full, continuum-mechanics-based approach is certainly applicable to test mass suspension systems in gravitational wave interferometers, it is often more useful to describe such a system with only a finite number of degrees of freedom. In that case it is not possible to assign one temperature to each degree of freedom. Instead we need to split up the system's impedance matrix $\bf{Z}$, relating the velocity vector $\bf{v}$ and force vector $\bf{F}$ via
\begin{equation}
\label{eq:impedanceDef}
{\bf Z}{\bf v} ={\bf F},
\end{equation}
into individual pieces ${\bf Z}_l$ assigned 
to a single temperature $T_l$, as well as the free particle impedance ${\bf Z}_{free}$:
\begin{equation}
\label{eq:impedance}
{\bf Z}={\bf Z}_{free} + \sum_l  {\bf Z}_l. 
\end{equation}
That this is possible will become clear when expanding the system size to include the normally ignored local degrees of freedom, see section \ref{sec:derivation}.
The free particle impedance is an imaginary diagonal matrix of the form 
${\bf Z}_{free,kk} = i \omega m_k$, where $m_k$ is the effective mass of the $k^{th}$ degree of freedom, and hence we have ${\bf Z}_{free} + {\bf Z}^{\dagger}_{free} = 0$, meaning that it is non-dissipative ($^{\dagger}$ denotes the complex conjugate and transposed matrix).
With this definition the (single-sided) force power spectral density matrix becomes
\begin{equation}
\label{eq:MasterNoiseMechFiniteF}
S^1_{{\bf F} {\bf F}}(f) = {2 k_B} \sum_l T_l  \left({\bf Z}_l +  {\bf Z}^{\dagger}_l \right) ,
\end{equation}
while the (single-sided) displacement power spectral density matrix is given by
\begin{equation}
\label{eq:MasterNoiseMechFinite}
S^1_{{\bf x} {\bf x}}(f) = \frac{2 k_B}{\omega^2} \sum_l T_l  {\bf Z}^{-1} \left({\bf Z}_l +  {\bf Z}^{\dagger}_l \right) {{\bf Z}^{-1}}^{\dagger} ,
\end{equation}
The sum in Eq.\ (\ref{eq:MasterNoiseMechFinite}) is equivalent to the dissipation-density-weighted integral over the volume in Eq.\  (\ref{eq:MasterNoise}). 
It is worth noting that while Eqs.\ (\ref{eq:MasterNoise}) and (\ref{eq:MasterNoiseMech}) seem to imply that the displacement thermal noise power is linear in the loss parameter, Eq.\  (\ref{eq:MasterNoiseMechFinite}) makes it clear that a high amount of dissipation will change the system's response to the thermal force noise. This dependence is implicit in the definition of the dissipation density $w_{\rm diss}(\vec{r},f)$.
Finally, if all temperatures $T_l$ are identical, the sum in Eq.\ (\ref{eq:MasterNoiseMechFinite}) reduces to $T ({\bf Z}^{-1} + {{\bf Z}^{-1}}^{\dagger} )$,
which is familiar from the equilibrium form of the fluctuation dissipation theorem. The same simplification is not possible in the non-equilibrium case because fundamentally the force noise is local in origin.

\section{Limitations}
\label{LIM}
When a heat flow is present in a system, it is in principle possible to convert a significant fraction of that energy to large-scale mechanical motion - that is after all the definition of a heat engine. Maybe the best example is the Sterling engine. However, from a noise point of view, a heat engine is an instability in one degree of freedom of the mechanical system due to the presence of a heat flow. It can only occur because the mechanical motion in that degree of freedom can affect the heat flow, and the modulated heat flow in turn drives the mechanical degree of freedom. In other words, we need a reciprocal mechanism from the mechanical motion to the heat flow.

A suspension system in a gravitational-wave interferometer, to a very good approximation, is designed to avoid such feed-back to the heat flow. We thus will assume for this paper that the heat flow, and therefore the temperature profile, is stationary and independent of the mechanical state. However, should such a feed-back mechanism exist in the considered system - even when it is too weak to drive the system into instability -  it can be modeled as a modified equation of motion for the associated thermal degree of freedom.

An additional limitation is related to the assumption that the temperature is well-defined locally. This is commonly assumed for heat flow calculations in a technical apparatus, and means that we can find small, but finite volume elements which are in thermal equilibrium. 
The literature on non-equilibrium thermodynamics refers to this as local thermodynamic equilibrium (LTE), or near-equilibrium conditions. Under LTE conditions the different operational ways to define temperature (kinetic temperature, configurational temperature, etc) all agree locally \cite{PhysRevE.95.013302}. 

In practice LTE conditions imply that all degrees of freedom of the thermal bath relevant for one volume element have to be spatially localized. While this is typically a good assumption for the macroscopic suspension systems used in gravitational-wave interferometers, there are obvious exceptions. One example is interaction with a radiation field that can have a very different temperature from the local equilibrium temperature. In that situation we could still generalize the fluctuation-dissipation theorem by adding extra impedance terms to Eq.\  (\ref{eq:impedance}) and associating them with the radiation temperature.

\section{Derivation}
\label{sec:derivation}
Once we assume a stationary temperature field $T(\vec{r})$, each volume element of the mechanical system has a well defined temperature. Any sufficiently localized mechanical sub-system with no long-range interactions thus can be described using a single temperature and the traditional fluctuation-dissipation theorem. The key for deriving the non-equilibrium version of the fluctuation-dissipation theorem given by Eqs\ (\ref{eq:MasterNoise}) through (\ref{eq:MasterNoiseMechFinite}) is thus to expand the mechanical system's number of degrees of freedom in such a way that the interactions of all individual degrees of freedom  become sufficiently localized to be described by a single temperature. For this enlarged system we can then write down the impedance matrix, splitting off the dissipation-free part ${\bf Z}_{free}$, as
\begin{equation}
\label{eq:impedanceProof}
{\bf Z}={\bf Z}_{free} + \sum_l  {\bf Z}_l = {\bf Z}_{free} + \int {\bf z}(\vec{r}) d^3r . 
\end{equation}
In the sum the index $l$ runs over pairs of interacting degrees of freedom - typically physical neighbors. Note that we can express the impedance ${\bf Z}$ (but not its inverse ${\bf Z}^{-1}$) as sum of partial impedances because the individual interaction forces are additive.
Because the origin of those interactions are all localized, we can write this sum as an integral over the system's volume by introducing the impedance matrix density ${\bf z}(\vec{r})$. It describes all interactions in the system due to the volume element $d^3r$. To illustrate this idea, it can be helpful to look at the example of a simple spring (as we will in section \ref{spring}), where the volume integral in Eq.\ (\ref{eq:impedanceProof}) is an integral over the volume of the elastic material of the spring, or equivalently a sum over the infinitesimal sub-springs.

Each one of the volume elements $d^3r$ is now giving rise to an interaction, and thus also  produces a force thermal noise. Since we assume LTE conditions, this volume element has a unique temperature $T(\vec{r})$, there is no ambiguity as to what temperature we should use, and we can apply the thermal equilibrium form of the fluctuation-dissipation theorem.
Thus, the force power spectral density matrix due to the volume element $d^3r$, described by index $l$, is given by
\begin{equation}
\label{eq:ForceNoiseProof}
S^1_{{\bf F} {\bf F},l}(f) = {2 k_B}  T_l  \left({\bf Z}_l +  {\bf Z}^{\dagger}_l \right) ,
\end{equation}
or, using the integral notation, 
\begin{equation}
\label{eq:ForceNoiseProof}
s^1_{{\bf F} {\bf F}}(f,\vec{r})  d^3r
= {2 k_B}  T(\vec{r})  \left({\bf z}(\vec{r}) +  {\bf z}^{\dagger}(\vec{r}) \right) d^3r .
\end{equation}
Furthermore, because it originates from independent interactions, the force noise from each volume element is uncorrelated with the force noise from any other element. Thus we can sum up the force power spectral density matrices from all volume elements, resulting in Eq.\ (\ref{eq:MasterNoiseMechFiniteF}). Equation (\ref{eq:MasterNoiseMechFinite}) then follows by applying the mechanical system response (Eq.\ (\ref{eq:impedanceDef})) to the thermal driving force.

In our derivation of Eqs.\ (\ref{eq:MasterNoiseMechFiniteF}) and (\ref{eq:MasterNoiseMechFinite}) it was instructive to expand the system to include localized degrees of freedom. However, we should point out that all we needed for the proof was an expansion of the total impedance matrix in terms of contributions from regions of well defined temperature. In other word, we can also perform this calculation using a simplified set of global degrees of freedom such as for example the 3 positions and 3 angles of a test mass in a pendulum suspension. We then just need to write the impedance matrix of the system as a sum of terms origination from a region of constant temperature, and Eqs.\ (\ref{eq:MasterNoiseMechFiniteF}) and (\ref{eq:MasterNoiseMechFinite}) will remain valid.

To conclude the proof, we need to show that the integral form of Eq.\  (\ref{eq:MasterNoise}) also follows from Eq.\  (\ref{eq:MasterNoiseMechFinite}). We note that 
\begin{equation}
\label{eq:ProfOfMasterNoise1}
S^1_{x x}(f) = {\bf q}^T S^1_{{\bf x} {\bf x}}(f) {\bf q} ,
\end{equation}
where we wrote the readout weights ${q}(\vec{r})$ in discrete form as real-valued column vector ${\bf q}$, and therefore get
\begin{equation}
\label{eq:ProfOfMasterNoise2}
S^1_{x x}(f)
= \frac{2 k_B}{\omega^2} \int  T(\vec{r}) \left[ {\bf q}^T {\bf Z}^{-1} \left({\bf z}(\vec{r}) +  {\bf z}^{\dagger}(\vec{r}) \right) {{\bf Z}^{-1}}^{\dagger} {\bf q} \right]  d^3r ,
\end{equation}
where we used the integral form of Eq.\  (\ref{eq:MasterNoiseMechFinite}).  
The dissipation density $w_{\rm diss}(\vec{r},f)$ for a system moving with the velocity vector ${\bf v}$ is given by
\begin{equation}
\label{eq:dissipationDensityProof1}
w_{\rm diss}(\vec{r},f) = {\bf v}^\dagger (f) \left({\bf z}(\vec{r}) +  {\bf z}^{\dagger}(\vec{r}) \right) {\bf v} (f) .
\end{equation}
Using the definition of impedance, Eq.\ (\ref{eq:impedanceDef}), and driving the system
with a force ${\bf F}(t) = F_0 {\bf q} \cos(2 \pi f t))$ we find
\begin{equation}
\label{eq:dissipationDensityProof2}
w_{\rm diss}(\vec{r},f) = \frac{F_0^2}{4} \left[ {\bf q}^T {{\bf Z}^{-1}}^{\dagger} \left({\bf z}(\vec{r}) +  {\bf z}^{\dagger}(\vec{r}) \right) {{\bf Z}^{-1}} {\bf q} \right] . 
\end{equation}
We note that Newton's third law implies that all impedance matrices are symmetric, ${\bf z}^T(\vec{r})={\bf z}(\vec{r})$ and ${\bf Z}^T={\bf Z}$. Since ${\bf q}$ is real, this implies
\begin{equation}
\label{eq:dissipationDensityProof3}
w_{\rm diss}(\vec{r},f) = \frac{F_0^2}{4} \left[ {\bf q}^T {\bf Z}^{-1} \left({\bf z}(\vec{r}) +  {\bf z}^{\dagger}(\vec{r}) \right) {{\bf Z}^{-1}}^{\dagger} {\bf q} \right] ,
\end{equation}
which, together with Eq.\ (\ref{eq:ProfOfMasterNoise2}), proves Eq.\  (\ref{eq:MasterNoise}).

\section{Application to a simple spring}
\label{spring}

First we discuss the thermal noise of a simple spring with a non-constant temperature profile attached to a test particle.
The spring is divided to n pieces and each piece except for the test particle has the small mass of $m_l,\,l=1,...,n-1$. We label the total spring constant $K$ and the test particle mass $m_n=M (\gg m_l)$.
At every point $l=1,...,n$ we define displacement, spring constant, and temperature as $x_l, {\bar k}_l,$ and $T_l$.
The spring constant is a complex value and consists of the real value $k_l$ and the loss angle $\phi_l$ as ${\bar k}_l=k_l (1+i\phi_l)$.

Total potential energy of the whole system $V_{\rm total}$ is given by
\begin{equation}
\label{eq:potentialspring}
V_{\rm total} = \frac{1}{2} \sum_l k_l ( x_l - x_{l-1} )^2,
\end{equation}
where $x_0=0$.
The equation of motion of each piece is given by
\begin{equation}
\label{eq:eqofmotion1}
-m_l \omega^2 x_l = -\cfrac{\partial V_{\rm total}}{\partial x_l} + F_l,
\end{equation}
where $F_l$ is the external force added to $m_l$, specifically
\begin{align}
\label{eq:eqofmotion2}
\begin{cases}
-m_1\omega^2 x_1 + {\bar k}_1 x_1 + {\bar k}_2 (x_1-x_2) &= F_1 \\
-m_l\omega^2 x_l + {\bar k}_l (x_l-x_{l-1}) - {\bar k}_{l+1} (x_{l+1}-x_l) &= F_l \\
-M\omega^2 x_n + {\bar k}_n (x_n-x_{n-1}) &= F_n,
\end{cases}
\end{align}
where we have written down the cases $l=1$ and $l=n$ explicitly, and take $l=2,...,n-1$ for the middle equation.
We now set the small mass of the spring $m$ to zero, which eliminates the internal excitation degrees of freedom by moving them to an infinitely large frequency. We then get the impedance of the system as
\begin{equation}
\label{eq:impedanceofspring}
i\omega {\bf Z} = \left(
\begin{array}{ccccc}
{\bar k}_1+{\bar k}_2 & -{\bar k}_2 & \ & \ & \ \\
-{\bar k}_2 & {\bar k}_2+{\bar k}_3 & -{\bar k}_3 & \ & \ \\
\ & -{\bar k}_3 & {\bar k}_3+{\bar k}_4 & \ddots & \ \\
\ & \ & \ddots & \ddots & -{\bar k}_{n} \\
\ & \ & \ & -{\bar k}_{n} & {\bar k}_n - M\omega^2
\end{array}
\right),
\end{equation}
where a blank implies that the matrix element is zero. Since each individual spring element $l$ has a unique temperature $T_l$, this equation not only describes the full impedance matrix ${\bf Z}$ , but also splits it up into a sum of individual pieces, with each piece being associated with a unique temperature, as required by Eq.\  (\ref{eq:impedance}).

Assuming that the spring constant and the loss angle of all pieces are the same, $k_l=k, \phi_l=\phi$, the force power spectral density matrix can be calculated using Eq.\  (\ref{eq:MasterNoiseMechFiniteF}) as
\begin{equation}
\label{eq:coilspringsff}
S^1_{{\bf F} {\bf F}}(f) = \cfrac{4 k_B k\phi}{\omega} \left(
\begin{array}{ccccc}
T_1+T_2 & -T_2 & \ & \ & \ \\
-T_2 & T_2+T_3 & -T_3 & \ & \ \\
\ & -T_3 & T_3+T_4 & \ddots & \ \\
\ & \ & \ddots & \ddots & -T_{n} \\
\ & \ & \ & -T_{n} & T_n
\end{array}
\right).
\end{equation}
Describing the inverse of the impedance matrix in terms of row vectors $ {\boldsymbol \zeta}^T_l$ as ${\bf Z}^{-1} = ( {\boldsymbol \zeta}^T_1;\ {\boldsymbol \zeta}^T_2;\ {\boldsymbol \zeta}^T_3;\ \cdots\ ;{\boldsymbol \zeta}^T_n)$, the displacement spectral density of the last (n-th) piece, i.e. the test particle,  can be written as
\begin{equation}
\label{eq:MasterNoiseMechFinitenn}
S^1_{nn}(f) = \frac{2 k_B}{\omega^2} {\boldsymbol \zeta}^{T}_n \sum_l T_l \left({\bf Z}_l +  {\bf Z}^{\dagger}_l \right) {\boldsymbol \zeta}_n^*,
\end{equation}
where the last (n-th) row vector is given by
\begin{equation}
\label{eq:inversenthcomponents}
{\boldsymbol \zeta}^{T}_n = \frac{i\omega}{k(1+i\phi) - nM\omega^2}
\left(
\begin{array}{ccccc}
1 & 2 & 3 & \cdots & n
\end{array}
\right).
\end{equation}
Noting that $k=nK$, the displacement spectral density can be calculated as
\begin{equation}
S^1_{nn}(f) = \frac{4k_B}{\omega n} \sum_l T_l \frac{K\phi}{(K-M\omega^2)^2 + K^2 \phi^2}.
\end{equation}
This result means that the average temperature of the whole system contributes to the displacement of thermal noise. Since the dissipation in this example is uniform across the spring, this result is expected based on equation \ref{eq:MasterNoise}.

\section{Application to a suspension fiber}
\label{fiber}
Next, we calculate the thermal noise of a suspension fiber.
As with the case of a simple spring, the suspension fiber is divided to $n$ pieces and each $n-1$ piece and the $n$-th mass has the mass of $m$ and $M$.
The angle of $l$-th piece against vertical direction is defined by $\theta_l \equiv (x_l - x_{l-1})/{\Delta z}$, where $x_l$ is the displacement of the $l$-th fiber along horizontal axis and ${\Delta z}$ is the length of the $l$-th fiber.
Total potential energy of that case can be written as
\begin{equation}
\label{eq:potentialfiber}
V_{\rm total} = \sum_{l=1}^{n+1} \cfrac{m_l g {\Delta z}}{2} \sum_{k=1}^{l} \theta_l^2 + \sum_{l=1}^{n+1} \cfrac{{\bar E}_l I}{2 {\Delta z}} (\theta_l - \theta_{l-1})^2,
\end{equation}
where ${\bar E}_l \equiv E_l(1+i\phi_l)$ is the complex Young's modulus of the fiber, $I=\int x^2 dA$ is the area moment of inertia in the direction of the horizontal axis, and $g$ gravity acceleration, respectively.
The first term derives from the gravity potential of each piece and the second term derives from the elastic energy of each fiber.
We set the boundary condition of $\theta_0 = \theta_{n+1}=0,\ \theta_1=x_1/{\Delta z}$.
In words, the upper clamp point is fixed, and the fiber is completely vertical at the upper and lower clamp points. While other boundary conditions are possible for a single fibers, this choice is required for 
the case of four-fiber suspensions as in the case of KAGRA.

The total impedance of the system can be calculated from the equations of motion as in the case of a simple spring.
Here we again assume that the Young's modulus of the fiber $E_l$, and in particular it's loss angle $\phi_l$, are independent of the position along the fiber.
Using our boundary conditions, the equations of motion on the 1-st, 2-nd, a generic $i$-th, $(n-1)$-th, and $n$-th piece are given by 
\begin{align}
\label{eq:eqofmotionboundary}
\begin{cases}
-m\omega^2 x_1 + \cfrac{Mg}{\Delta z} \left( 2x_1 - x_2 \right) \\ + \cfrac{E(1+i\phi) I}{{\Delta z}^3} \left( 6x_1-4x_2+x_3 \right) &= F_1 \\
-m\omega^2 x_2 + \cfrac{Mg}{\Delta z} \left( -x_1 + 2x_2 - x_3 \right) \\ + \cfrac{E(1+i\phi) I}{{\Delta z}^3} \left( -4x_1 + 6x_2 -4x_3 + x_4 \right) &= F_2 \\
-m\omega^2 x_i + \cfrac{Mg}{\Delta z} \left( -x_{i-1} + 2x_i - x_{i+1} \right) \\ + \cfrac{E(1+i\phi) I}{{\Delta z}^3} \left( x_{i-2} -4x_{i-1} + 6x_i + 4x_{i+1} + x_i \right) &= F_i \\
-m\omega^2 x_{n-1} + \cfrac{Mg}{\Delta z} \left( -x_{n-2}+2x_{n-1} - x_n \right) \\ + \cfrac{E(1+i\phi) I}{{\Delta z}^3} 
\left( x_{n-3}-4x_{n-2}+6x_{n-1}-3x_n \right) &= F_{n-1} \\
-M\omega^2 x_n + \cfrac{Mg}{\Delta z} \left( -x_{n-1} + x_n \right) \\ + \cfrac{E(1+i\phi) I}{{\Delta z}^3} \left( x_{n-2}-3x_{n-1}+2x_n \right) &= F_n.
\end{cases}
\end{align}

Dividing total impedance into three parts
\begin{equation}
\label{eq:totalimpedance}
{\bf Z}_{total} = {\bf Z}_{free} + {\bf Z}_{grav} + {\bf Z}_{elas},
\end{equation}
they can be written as 
\begin{equation}
\label{eq:threeimpedance1}
i\omega {\bf Z}_{free} = \left(
\begin{array}{ccccc}
-m\omega^2 & \ & \ & \ & \ \\
\ & -m\omega^2 & \ & \ & \ \\
\ & \ & -m\omega^2 & \ & \ \\
\ & \ & \ & \ddots & \ \\
\ & \ & \ & \ & -M\omega^2
\end{array}
\right),
\end{equation}
\begin{equation}
\label{eq:threeimpedance2}
i\omega {\bf Z}_{grav} \simeq \frac{Mg}{{\Delta z}} \left(
\begin{array}{ccccc}
2 & -1 & \ & \ & \ \\
-1 & 2 & -1 & \ & \ \\
\ & -1 & 2 & \ddots & \ \\
\ & \ & \ddots & \ddots & -1 \\
\ & \ & \ & -1 & 1
\end{array}
\right),
\end{equation}
\begin{equation}
\label{eq:threeimpedance3}
i\omega {\bf Z}_{elas} = \frac{E(1+i\phi) I}{{\Delta z}^3} \left(
\begin{array}{ccccc}
6 & -4 & 1 & \ & \ \\
-4 & 6 & -4 & \ddots & \ \\
1 & -4 & 6 & \ddots & 1 \\
\ & \ddots & \ddots & \ddots & -3 \\
\ & \ & 1 & -3 & 2
\end{array}
\right),
\end{equation}
Here we do not consider $m$ in ${\bf Z}_{free}$ to be zero in order to recover the violin modes of the fiber. Note that we could choose not to set the loss angles $\phi_l$ of all fiber pieces to be the same. In that case Eq.\ (\ref{eq:threeimpedance3}) becomes a sum of matrices  over individual segments $l$, each with its own associated temperature $T_l$. This is similar to Eq.\ (\ref{eq:coilspringsff}), and as expected based on Eq.\ (\ref{eq:impedance}).

\begin{figure}
\centering
\includegraphics[width=\hsize]{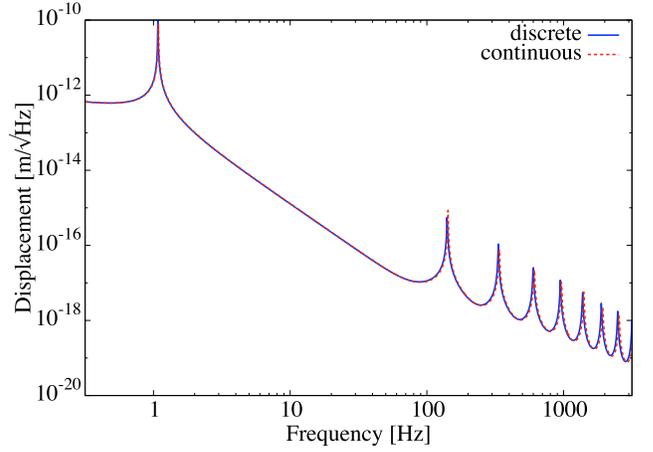}
\caption{
Discrete and continuous calculation of the suspension thermal noise of a fiber. We set $m=1$ kg, the length of the fiber $L=30$ cm, the radius of the fiber $r=0.5$ mm, $\phi = 1 \times 10^{-2}$, and $T=300$ K. The material of the fiber is assumed to be sapphire. In the discrete calculation the fiber is divided to 100 pieces.
}
\label{fig:ananumcompare}
\end{figure}

With this total impedance, we can calculate the thermal noise of a suspension fiber numerically. First of all, we demonstrate the validity of the calculation with the simple situation.
Assuming that the temperature is constant, we compare the displacement thermal noise from discrete calculation of \ref{eq:MasterNoiseMechFinite} and continuous calculation of~\cite{PPPnote}.
The result is shown in Fig. \ref{fig:ananumcompare}.
The floor level of the noise, the resonant frequency of the pendulum mode and violin modes are different by less than 1\%, around 1\% and 2\%, respectively.
This value is reasonable because the number of divided pieces is $n=100$ and the precision should be the order of $1/n$.
The first peak around $1~{\rm Hz}$ is the pendulum mode, while the peaks at $140~{\rm Hz}$ and harmonics are the violin modes of the fiber.

\begin{figure}
\centering
\includegraphics[width=\hsize]{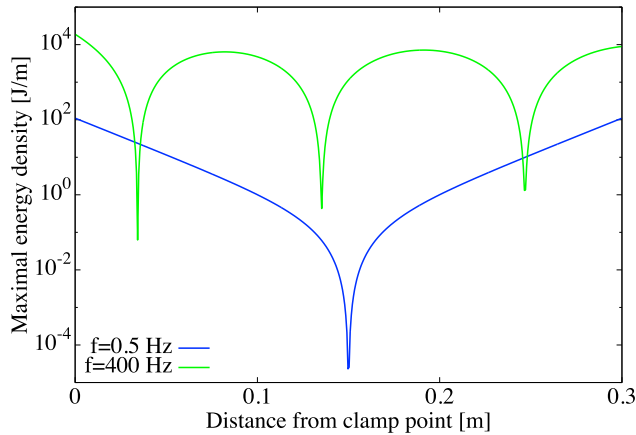}
\caption{
Elastic energy distribution along the suspension fiber for two frequencies, 0.5 Hz and 400 Hz. The distributions are calculated using the continuos model derived in \cite{PPPnote}, but agree with our discrete model. The two traces illustrate that different frequencies sample the temperature and mechanical loss at different locations along the fiber, in accordance with Eq.\ (\ref{eq:MasterNoiseMech}).
}
\label{fig:elasticenergy}
\end{figure}

Next we look at non-uniform temperature distributions. To get an intuitive understanding of the physics involved, we start with plotting the elastic energy distribution in Fig. \ref{fig:elasticenergy} for two examples: i) A frequency below the pendulum mode frequency. The fiber is mostly bending near the clamp point and the test mass attachment point, while the center of the fiber is not deformed. And ii) a frequency between violin modes. The dips correspond to nodes of the induced motion, where again the fiber is not deformed.
The traces are calculated using the continuous model derived in 
\cite{PPPnote}, which describes the elastic energy distribution along the position of the fiber, 
but agree with our discrete model.

\begin{figure}
\centering
\includegraphics[width=\hsize]{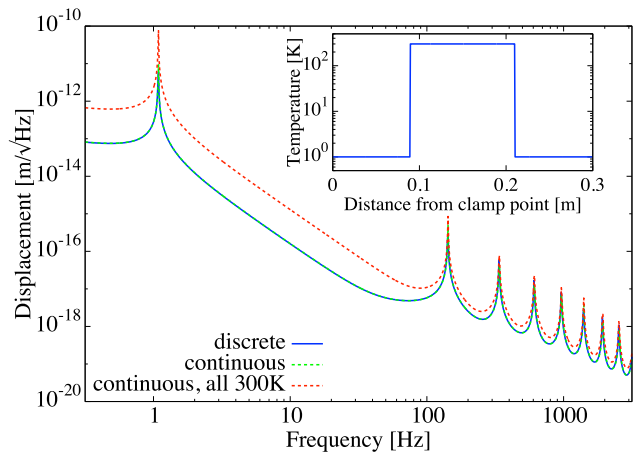}
\caption{
Discrete and continuous calculation of the suspension thermal noise with a non-uniform temperature distribution. To highlight the effect on the thermal noise power spectrum we chose the extreme temperature distribution shown in the inset, with non-zero temperature only in the middle third section of the fiber. The blue solid line shows the result of the discrete calculation. The green dotted line shows the continuous result. For reference, the red dotted line shows the same noise as in Fig.\ \ref{fig:ananumcompare}, i.e. for a uniform temperature everywhere along the fiber.
The parameters are all the same as for the previous simulation, except for the number of discrete fiber sections, which is set to $n=500$.
}
\label{fig:tempdistthermal}
\end{figure}

To demonstrate the effect of non-uniform temperature distributions we start with an extreme, although unphysical example. We assume an elevated temperature ($300~\rm{K}$) for only the middle section of the fiber, as illustrated in the inset of Fig.\ \ref{fig:tempdistthermal}.
The main part of that figure shows the thermal noise for this temperature distribution, calculated using our discrete model (solid blue), as well using the continuous model (dotted green). For reference the figure also shows the thermal noise for a uniform temperature of $300~\rm{K}$ along the fiber (dotted red). This is the same trace as in Fig.\ \ref{fig:ananumcompare}.
Compared to this red trace, the noise level at low frequencies is improved significantly because the energy loss of the pendulum mode comes from the large distortion around the clamp point and attachment point, where temperature is much lower than that at the center. Compared with low frequencies, the noise of violin modes does not change significantly because some of the antinodes of the energy distribution profile lie in the $300~\rm{K}$ region.
Finally, the blue and green traces agree within the numerical uncertainties, validating our discrete model and Eq. \ref{eq:MasterNoise}.

\section{The KAGRA suspension thermal noise}
\label{KAGRA}
The main test masses of KAGRA are suspended by an eight-stage pendulum called Type-A system. 
The last four-stage payload of the Type-A system is cooled down to cryogenic temperature and is called a cryopayload~\cite{1742-6596-716-1-012017}. 
Here we calculate the thermal noise of the KAGRA cryopayload for the input test mass (ITM). Brownian thermal noise is considered since it is dominant as compared with thermo-elastic noise.

\begin{figure}
\centering
\includegraphics[width=\hsize]{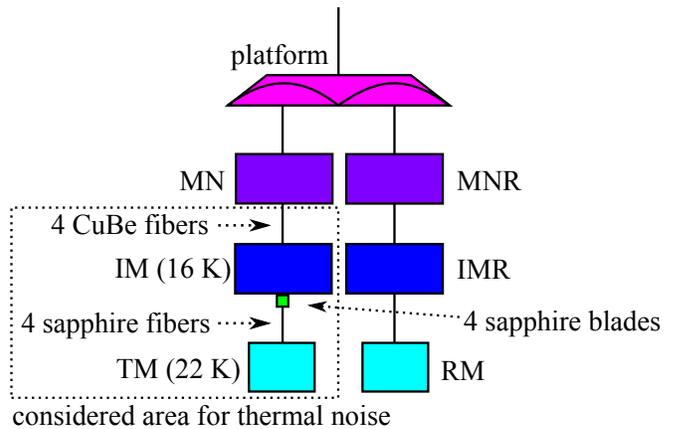}
\caption{
Schematic of the KAGRA cryopayload.
Suspension thermal noise on KAGRA derives from the surrounded area by dotted lines.
}
\label{fig:Cryopaylaod}
\end{figure}

Figure~\ref{fig:Cryopaylaod} shows the schematic of the KAGRA cryopayload. 
The platform is suspended from upper room temperature stages. 
The marionette is suspended from the platform with 1 maraging steel fiber. 
The intermediate mass (IM, 20.8 kg) is suspended from the marionette with 4 copper beryllium (CuBe) fibers (26.1 cm long, 0.6 mm dia.). 
Finally, the sapphire test mass (TM, 22.7 kg) is suspended from the 4 sapphire blades (0.1 kg) attached to the intermediate mass with 4 sapphire fibers (35 cm long, 1.6 mm dia.).

\begin{figure}
\centering
\includegraphics[width=\hsize]{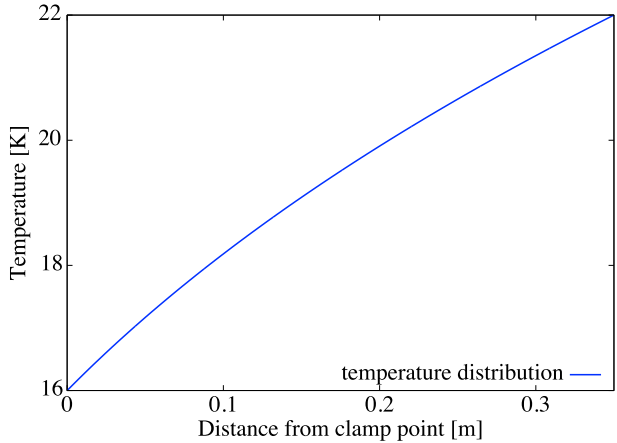}
\caption{
Temperature distribution of the KAGRA sapphire fibers suspending the test mass. This is derived by solving a differential equation about the temperature function of the cramp point. The boundary condition is $T(z=0)=16$ K, and $T(z=0.35)=22$ K.
}
\label{fig:temperatureprofile}
\end{figure}

Aluminum heat links are attached to the marionette and the marionette is cooled down to 15 K. 
Heat absorption of the laser beam and the thermal conductivity of the fibers define the test mass temperature. 
Estimated temperature profile along the sapphire fiber is plotted in Fig.~\ref{fig:temperatureprofile}. 
Here we assumed temperature of the IM and the TM to be 16~K and 22~K, respectively, the incident beam power from the back surface of the input test mass to be 674~W, mirror substrate absorption to be 50~ppm/cm, and coating absorption to be 0.5~ppm. This results in a nominal power loading of $0.724~\rm{W}$ for the input test mass.
We used the measured thermal conductivity of the sapphire fiber in Ref.~\cite{0264-9381-31-10-105004}, $\kappa (T) = 7.98\times T^{2.2}$~W/K/m.

We discuss horizontal suspension thermal noise including the system below the CuBe fibers.
It is enough to consider a pendulum mode at the second pendulum consisting of the CuBe fibers and the IM because displacement thermal noise deriving from the second pendulum has dependence of $f^{-4.5}$ above resonant frequency of the differential pendulum mode (1.9~Hz) resulting in that violin modes of CuBe fibers are negligible.
Therefore, 4 CuBe fibers can be regarded as effective one fiber, whose tension is 1/4 but the spring constant is 4 times larger.
The horizontal potential is written as
\begin{align}
\label{eq:potentialkagrahor}
V_{\rm hor} = \cfrac{1}{2}k_{{\rm IM},h}x_{\rm IM}^2 + \sum_{\mu=a,b,c,d} \left[ \cfrac{1}{2}k_{{\rm bl},h,\mu} (x_{{\rm bl},\mu}-x_{\rm IM})^2 \right. \nonumber \\ 
\left. + \sum_{l=1}^{n+1} \cfrac{m_l g {\Delta z}}{2} \sum_{k=1}^{l} \theta_{l,\mu}^2 + \sum_{l=1}^{n+1} \cfrac{{\bar E}_{l,\mu} I}{2 {\Delta z}} (\theta_{l,\mu} - \theta_{l-1,\mu})^2 \right],
\end{align}
where $k_{\rm IM,h}, k_{{\rm bl},h,\mu}, x_{\rm IM}, x_{{\rm bl}, \mu}$ are the horizontal spring constant and the displacement of the IM and blade springs.
The labeling of $a,b,c,d$ means 4 blade springs and 4 sapphire fibers.
The boundary condition is $\theta_{0,\mu}=\theta_{n+1,\mu}=0$ and $\theta_{1,\mu}=(x_{1,\mu}-x_{{\rm bl}, \mu})/{\Delta z}$.
We can get full horizontal thermal noise by doing the same numerical calculation with this potential.

\begin{figure}
\centering
\includegraphics[width=\hsize]{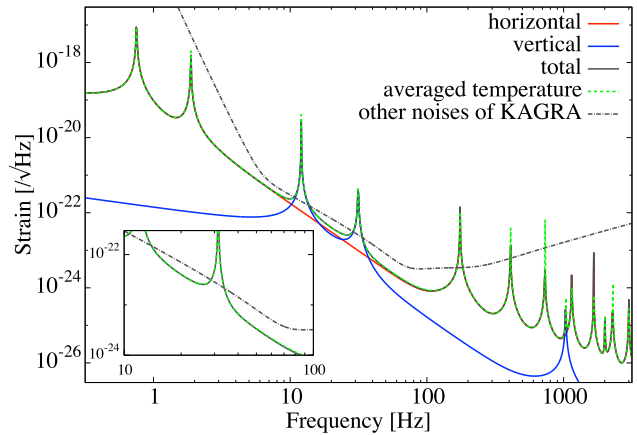}
\caption{
Total suspension thermal noise of KAGRA in strain considering the temperature distribution. The arm length of KAGRA is 3 km. Horizontal and vertical resonant frequency of the blade spring is assumed to be 2 kHz and 14.5 Hz (with suspending the test mass), respectively. First two peaks come from the common and differential pendulum mode. The peak around 30 Hz is due to resonance of CuBe fiber bounce. The resonant frequency of first violin mode is around 180 Hz. The green dotted line shows suspension thermal noise using an averaged temperature of the IM and TM. The dashed gray line shows the level of other expected noise sources in KAGRA.
}
\label{fig:kagrathermalnoise}
\end{figure}

Subsequently we consider the vertical thermal noise below CuBe fibers similarly.
A spring constant of a vertical bounce mode can be described as $k_v=ES/L$, where $E$ is the Young's modulus, $S$ is the surface area, and $L$ is the length of the fiber.
Thus, 4 fibers can be regarded as one fiber by taking 4 times surface.
The vertical potential is written as
\begin{align}
\label{eq:potentialkagraver}
V_{\rm ver} = \cfrac{1}{2}k_{{\rm IM},v}x_{\rm IM}^2 + \cfrac{1}{2}k_{{\rm bl},v} (x_{\rm bl}-x_{\rm IM})^2 \nonumber \\ 
+\cfrac{1}{2} \sum_{l=1}^{n} \cfrac{\bar{E}_l S}{{\Delta z}} (x_{l}-x_{l-1})^2,
\end{align}
where $k_{\rm IM,h}, k_{{\rm bl},v}$ is the vertical spring constant of the CuBe fibers and blade springs and $x_0=x_{\rm bl}$.
These two suspension thermal noises are shown in Fig. \ref{fig:kagrathermalnoise}.

In this figure we also compare the full numerical result to a simplified suspension thermal noise calculation that uses the average temperature of the IM and TM. The noise level between the two only differs by around 2\% in the frequency band of about 10 Hz to 50 Hz, where suspension thermal noise contributes most to the total noise. This result can be intuitively understood because the elastic energy is symmetric and the upper and lower edge of the fibers provide the largest contributions.  We thus conclude that for the practical purpose of predicting KAGRA's suspension thermal noise it is enough to
average the IM and TM temperature and use the equilibrium formulation of the fluctuation-dissipation theorem.

\section{Application to Thermo-Elastic and Thermo-Refractive Noise}
\label{sec:THELN}
The non-equilibrium fluctuation-dissipation theorem discussed above is also applicable to thermo-elastic and thermo-refractive noise. However we now encounter the complication that we need to calculate temperature fluctuations in the presence of a temperature gradient, which can proof challenging in practice. Nevertheless we can outline the guiding principles here.

To calculate thermo-elastic and thermo-refractive noise in a degree of freedom $x=\int {q}(\vec{r}){T}(\vec{r}) d^3r $, with ${q}(\vec{r})$ the readout weights for the temperature field ${T}(\vec{r})$, the procedure described by Levin \cite{LEVIN20081941} calls for driving the system with the entropy density $ds (\vec{r},f) = F_0 {q}(\vec{r}) \cos(2 \pi f t)$, and calculating the power dissipated in the system by thermal diffusion. Ignoring surface terms for simplicity, the time derivative of the entropy density is given by (see e.g. \cite{PhysRevD.90.043013})
\begin{equation}
\label{eq:TELdiss1}
\dot{s} = \vec{j} \cdot \vec{\nabla}\frac{1}{T} 
=- \frac{\vec{j} \cdot \vec{\nabla} \ln{T}}{T} \,\, .
\end{equation}
Thus the dissipation rate density $\dot{\mathpzc q}$ due to diffusion is given by
\begin{equation}
\label{eq:TELdiss2}
\dot{\mathpzc q} = T \dot{s} =
- \vec{j} \cdot \vec{\nabla} \ln{T} \,\, .
\end{equation}
If we want to use this expression in non-equilibrium conditions, there are two complications: i) Since the background temperature field can vary significantly, calculating the heat flow through the linearized expression $\vec{j}=-\kappa \vec{\nabla} T$ might not be adequate for the whole system. This is why we intentionally avoided introducing the thermal conductivity $\kappa$ in the equations \ref{eq:TELdiss1} and \ref{eq:TELdiss2}. ii) The expression in equation \ref{eq:TELdiss2} is non-zero for the background heat flow since setting up a stationary temperature gradient necessarily introduces stationary thermal dissipation. We can address item ii) by splitting heat flow and temperature field into a stationary zeroth-order term, $\vec{j}_0$ and ${T}_0$, and terms higher-order in the drive amplitude $F_0$.
To find the dissipation relevant for the fluctuation-dissipation theorem, we can then subtract the stationary background dissipation:
\begin{equation}
\label{eq:TELdiss4}
w_{\rm diss}(\vec{r},f) = \left<  \dot{\mathpzc q} - \dot{\mathpzc q}_0 \right> 
 \,\, .
\end{equation}
Here the $\left< ... \right>$ denotes cycle-averaging over one drive cycle. Note that the linear terms in the drive amplitude $F_0$ average to zero after one cycle, i.e. the dissipation $w_{\rm diss}$ is quadratic in $F_0$, as required by Eq.\ (\ref{eq:MasterNoise}). This expression for the dissipation density $w_{\rm diss}$ can then be used in Eq.\ (\ref{eq:MasterNoise}) to find the thermo-elastic and/or thermo-refractive noise in the degree of freedom $x$.

\section{Conclusion}
We expanded the application of the Fluctuation-Dissipation theorem for mechanical systems to non-equilibrium steady-state conditions in which the temperature is only defined locally. We note that the requirement of a stationary background temperature field rules out any feed-back from the mechanical motion to the heat flow, which is what would occur in a heat engine. To calculate the thermal noise, the correct weight for averaging the temperature field is given by the dissipation density of the mechanical system. For illustration purposes we apply this result to a simple spring and a fiber suspension, as well as to a model of the KAGRA gravitational-wave interferometer suspension. 

\section*{Acknowledgements}
We would like to acknowledge J. Harms, L. Conti, K. Shibata, and S. Ren for many fruitful discussions. This work was supported by JSPS KAKENHI Grant No. 16J01010 and NSF awards PHY1707876 and PHY1352511.

\bibliography{paper}

\begin{thebibliography}{10}

\bibitem{PhysRevD.88.043007}
The KAGRA Collaboration, Y.~Aso {\em et~al.},
\newblock Phys. Rev. D {\bf 88}, 043007 (2013).

\bibitem{0264-9381-32-2-024001}
F.~Acernese {\em et~al.},
\newblock Classical and Quantum Gravity {\bf 32}, 024001 (2015).

\bibitem{PhysRevLett.116.131103}
LIGO Scientific Collaboration and Virgo Collaboration, B.~P. Abbott {\em
  et~al.},
\newblock Phys. Rev. Lett. {\bf 116}, 131103 (2016).

\bibitem{Harry:06}
G.~M. Harry {\em et~al.},
\newblock Appl. Opt. {\bf 45}, 1569 (2006).

\bibitem{MATICHARD2015273}
F.~Matichard {\em et~al.},
\newblock Precision Engineering {\bf 40}, 273  (2015).

\bibitem{MATICHARD2015287}
F.~Matichard {\em et~al.},
\newblock Precision Engineering {\bf 40}, 287  (2015).

\bibitem{BRACCINI2005557}
S.~Braccini {\em et~al.},
\newblock Astroparticle Physics {\bf 23}, 557  (2005).

\bibitem{ACERNESE2010182}
F.~Acernese {\em et~al.},
\newblock Astroparticle Physics {\bf 33}, 182  (2010).

\bibitem{PhysRev.83.34}
H.~B. Callen and T.~A. Welton,
\newblock Phys. Rev. {\bf 83}, 34 (1951).

\bibitem{0034-4885-29-1-306}
R.~Kubo,
\newblock Reports on Progress in Physics {\bf 29}, 255 (1966).

\bibitem{PhysRevD.42.2437}
P.~R. Saulson,
\newblock Phys. Rev. D {\bf 42}, 2437 (1990).

\bibitem{PhysRevD.57.659}
Y.~Levin,
\newblock Phys. Rev. D {\bf 57}, 659 (1998).

\bibitem{LEVIN20081941}
Y.~Levin,
\newblock Physics Letters A {\bf 372}, 1941  (2008).

\bibitem{PhysRevD.78.102003}
M.~Evans {\em et~al.},
\newblock Phys. Rev. D {\bf 78}, 102003 (2008).

\bibitem{PhysRevD.91.023010}
S.~W. Ballmer,
\newblock Phys. Rev. D {\bf 91}, 023010 (2015).

\bibitem{ETDesign}
M.~Abernathy {\em et~al.},
\newblock Einstein gravitational wave {T}elescope conceptual design study,
\newblock ET-0106C-10, 2011.

\bibitem{0264-9381-27-19-194002}
M.~Punturo {\em et~al.},
\newblock Classical and Quantum Gravity {\bf 27}, 194002 (2010).

\bibitem{PhysRevD.91.082001}
S.~Dwyer {\em et~al.},
\newblock Phys. Rev. D {\bf 91}, 082001 (2015).

\bibitem{0264-9381-34-4-044001}
B.~P. Abbott {\em et~al.},
\newblock Classical and Quantum Gravity {\bf 34}, 044001 (2017).

\bibitem{PhysRevE.95.013302}
P.~K. Patra and R.~C. Batra,
\newblock Phys. Rev. E {\bf 95}, 013302 (2017).

\bibitem{PPPnote}
P.~P. Piergiovanni~F., Punturo~M.,
\newblock VIR-015E-09  (2009).

\bibitem{1742-6596-716-1-012017}
R.~Kumar {\em et~al.},
\newblock Journal of Physics: Conference Series {\bf 716}, 012017 (2016).

\bibitem{0264-9381-31-10-105004}
A.~Khalaidovski {\em et~al.},
\newblock Classical and Quantum Gravity {\bf 31}, 105004 (2014).

\bibitem{PhysRevD.90.043013}
S.~Dwyer and S.~W. Ballmer,
\newblock Phys. Rev. D {\bf 90}, 043013 (2014).

\end{thebibliography}
\bibliographystyle{h-physrev3}
\end{document}